\documentclass[aps,prb,reprint,superscriptaddress,secnumarabic,amssymb,nobibnotes,longbibliography]{revtex4-2}
\usepackage{graphicx}
\usepackage{amsmath}
\usepackage{subfigure}
\usepackage{verbatim}
\usepackage[utf8]{inputenc}
\usepackage{dsfont}
\usepackage[colorinlistoftodos]{todonotes}
\usepackage[version=3]{mhchem}
\usepackage{chemfig}
\usepackage{cancel}
\usepackage{hyperref}
\usepackage{xcolor}
\usepackage{upgreek}
\usepackage{bm}
\usepackage{amssymb}
\usepackage{braket}

\begin{document}

\title{Nonreciprocal ballistic transport in multi-layer Weyl semimetal films \\ with surface engineering}

\author{M. H. Zou}
\affiliation{National Laboratory of Solid State Microstructures, School of Physics,
and Collaborative Innovation Center of Advanced Microstructures, Nanjing University, Nanjing 210093, China}

\author{R. Ma}
\affiliation{
Nanjing University of Information Science and Technology, Nanjing 210044, China}

\author{S. J. Xu}
\affiliation{College of Physics, Nanjing University of Aeronautics and Astronautics, Nanjing 211106, China}

\author{W. Chen}
\affiliation{National Laboratory of Solid State Microstructures, School of Physics,
and Collaborative Innovation Center of Advanced Microstructures, Nanjing University, Nanjing 210093, China}

\author{H. Geng}
\email{genghao@nuaa.edu.cn}
\affiliation{College of Physics, Nanjing University of Aeronautics and Astronautics, Nanjing 211106, China}
\affiliation{Key Laboratory of Aerospace Information Materials and Physics (NUAA), MIIT, Nanjing 211106, China}

\author{L. Sheng}
\email{shengli@nju.edu.cn}
\affiliation{National Laboratory of Solid State Microstructures, School of Physics,
and Collaborative Innovation Center of Advanced Microstructures, Nanjing University, Nanjing 210093, China}

\author{D. Y. Xing}
\affiliation{National Laboratory of Solid State Microstructures, School of Physics,
and Collaborative Innovation Center of Advanced Microstructures, Nanjing University, Nanjing 210093, China}

\date{\today}

\begin{abstract}
    Weyl semimetal (WSM) thin films exhibit distinct electronic properties compared to their bulk counterparts. In this study, we theoretically investigate the nonreciprocal ballistic transport phenomena arising in WSM thin films due to surface modifications. Our analysis demonstrates that the nonreciprocity is sub-band-resolved, where the surface states provide the dominant contribution to the nonreciprocity, whereas the bulk states introduce a negative correction. Calculations further reveal a quantum size effect: overall, the nonreciprocal signal decreases with increasing film thickness, but it undergoes discontinuities as the Fermi energy approaches the bottom of a sub-band. Moreover, we observe that the density of states (DOS) in such multi-layer systems exhibits a thickness-independent pattern, which can be effectively explained by a single-variable theory.
\end{abstract}

\maketitle

\section{Introduction}

Nonreciprocity in charge transport reflects an asymmetric relationship between current and voltage, where the magnitude of current changes when the voltage has an opposite sign~\cite{Tokura2018, Ideue2021}.
Although nonreciprocal transport is commonly observed in systems breaking the translational symmetry, for example the p-n junction, the achievement of nonreciprocity in translationally symmetric systems has emerged more recently.
After extensive exploration to identify material candidates, there have been significant findings such as chiral nanosystems~\cite{Rikken2001, Krstić2002}, polar semiconductors~\cite{Ideue2017, Itahashi2020, Li2021}, bilayer heterojunctions~\cite{Avci2015, Yasuda2019, Choe2019, Shim2022, Ye2022}, and topological systems~\cite{Yasuda2020, Li2024}.
On the theoretical side, a variety of mechanisms have been proposed to explain the new phenomena, including asymmetric band structures~\cite{Tokura2018, Ideue2017}, asymmetric inelastic scattering by spin clusters~\cite{Ishizuka2020} and magnons~\cite{Yasuda2016}, quantum metric~\cite{Kaplan2024, Wang2023}, quantum interferences\cite{Mehraeen2023}, and the non-Hermitian skin effect~\cite{Geng2023, Shao2024}.
In this article, we concentrate on the nonreciprocal transport in systems characterized by asymmetric bands.

Compared to bulk materials which most previous theories focus on, the study of nonreciprocity in thin-film materials seems more valuable.
On the one hand, thin films are more commonly used in experimental measurements~\cite{Ideue2017}, some of which can already be fabricated to be quite thin~\cite{Kanagaraj2022, Leahy2024, Li2020a}.
On the other hand, the more pronounced quantum effects and surface effects allow thin films to possess different properties from their bulk samples.
The WSM thin film is a good example, which is a superior choice for the study of interfacial charge transfer effect~\cite{Wang2024, Nelson2022, Ostovar2024}, spin Hall effect~\cite{Sun2016, Zhao2020, Ng2021}, spin-charge pumping~\cite{Kurebayashi2016, Zhao2020}, spin-orbit torque effect~\cite{Kurebayashi2019, Bainsla2024, Sakamoto2025}, and finally thickness dependent magnetotransport measurements~\cite{Kanagaraj2022, Igarashi2017, Kaneta-Takada2021}.
Besides, compared to bulk samples, WSM thin films possess unique properties, such as anisotropic weak antilocalization~\cite{Leahy2024}, larger coercive fields, pronounced hysteresis in the magnetoresistance and highly tunable anomalous Hall conductivity~\cite{Li2020a}.
Recently, nonreciprocity induced by surface engineering has been predicted in WSM bulk systems~\cite{Jia2024}.
In order to meet the needs of actual experimental measurements, a theory of nonreciprocity in thin-film materials is necessary.

WSMs are three-dimensional topological quantum materials characterized by an even number of points in momentum space known as Weyl nodes (WNs)~\cite{Wan2011, Li2020}.
At certain surfaces of a WSM, topologically protected surface states known as Fermi arcs connect a couple of WNs of opposite chiralities.
Normally, either time-reversal symmetry (TRS) or spatial inversion symmetry (SIS) is maintained in the WSM, ensuring a symmetric band structure.
However, recent advancements in research suggest that Fermi arcs can be manipulated through surface modifications, making asymmetric band structures possible for WSMs~\cite{Morali2019, Ekahana2020, Souma2016, Sun2015, Zheng2023}, which would hopefully lead to nonreciprocity.
In this case, it is  the surface states (Fermi arcs) rather than the bulk states that dominate the nonreciprocity, which is rare in traditional materials.
Compared to bulk WSMs, such surface engineering is supposed to have a more pronounced impact to the films due to the greater surface effects.
Therefore, larger nonreciprocity is assumed to be achieved in WSM thin films.

\begin{figure}[t]
    \centering
    \includegraphics[width=0.75\columnwidth]{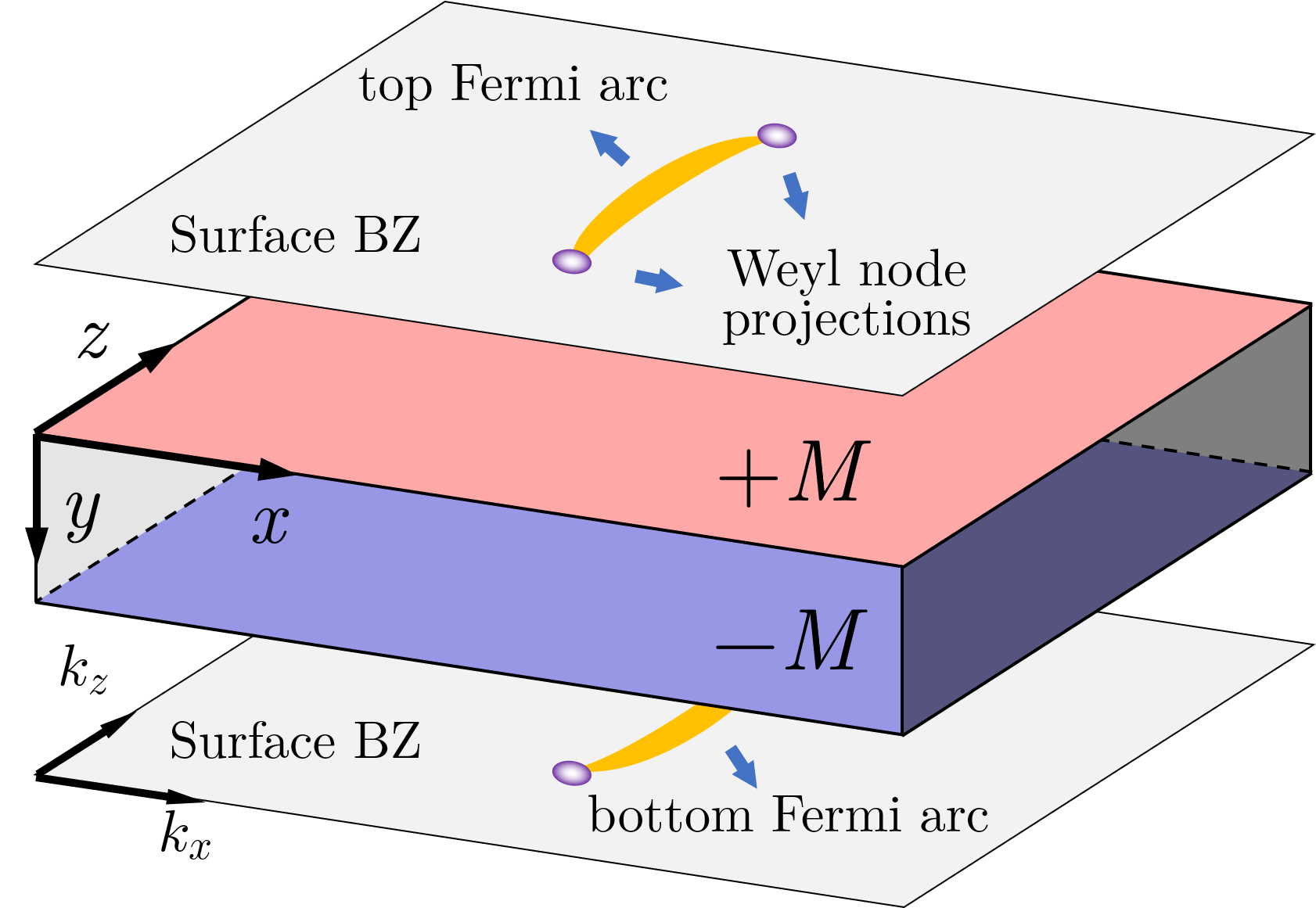}
    \caption{
        Schematic representation of a WSM film which is infinite in the $x-z$ plane but finite in the $y$ direction (middle), and its surface Brillouin zone (BZ) with WN projections connected by Fermi arcs on it (upper and lower).
        The two topologically nontrivial surfaces in red and blue are modified by the antisymmetric potentials $\pm M$.
    }\label{fig0}
\end{figure}

In this article, we calculate the nonreciprocity in the system of multi-layer WSM films.
Considering the significant quantum effects, the semiclassical Boltzmann transport theory is no longer applicable.
So we use the gauge-invariant theory in the ballistic regime~\cite{Zou2024}.
Besides easier fabrication and larger nonreciprocal signal in experiment, thin films make it possible to study how the surface states and bulk states contribute differently to nonreciprocity.
Furthermore, we also investigate the thickness dependence of the sample on the nonreciprocity.
Through this process, a thickness-independent pattern of the DOS is also observed.
Realizing its universality in multi-layer systems, we develop a single-variable theory to explain it.

The rest of the article is organized as follows.
In Sec.~\ref{sec. method}, the model Hamiltonian as well as the gauge-invariant theory for nonreciprocal ballistic transport is introduced.
In Sec.~\ref{sec. result}, we show the numerical results of the nonreciprocal signal and discuss the thickness-independent pattern of DOS.
Finally, in Sec.~\ref{sec. conclusion}, we present our concluding remarks.

\section{Model and Method}\label{sec. method}   

\subsection{Model Hamiltonian of the two-node WSM film with surface manipulation}
Consider the following Hamiltonian of a WSM on the simple cubic lattice as
\begin{align}
    H(\mathbf{k})
        &= t_1 [\sin (a_0 k_x) \sigma_x + \sin (a_0 k_y) \sigma_y]
            + m_\mathbf{k} \sigma_z, \nonumber\\
    m_\mathbf{k} &= t_2 [\cos (a_0 k_x) + \cos (a_0 k_y) + \cos (a_0 k_z) + m],
    \label{eq.ham_bulk2}
\end{align}
where $\mathbf{k} = (k_x, k_y, k_z)$ is the electron momentum;
$\sigma_x, \sigma_y, \sigma_z$ are the Pauli matrices;
$a_0$ is the lattice constant;
and $t_1, t_2, m$ denote model parameters.
In this model, TRS is broken while SIS is maintained.
Thus, there exit two Weyl nodes which are located separately at $(0, 0, \pm \arccos(-m-2))\,a_0^{-1}$.
The energy spectrum is
\begin{align}
    E_\mathbf{k} = \pm \sqrt{t_1^2 [\sin^2 (a_0 k_x) + \sin^2 (a_0 k_y)] + m_\mathbf{k}^2}.
\end{align}
The Weyl node energy is at $E=0$.

\begin{figure}[t]
    \centering
    \includegraphics[width=1\columnwidth]{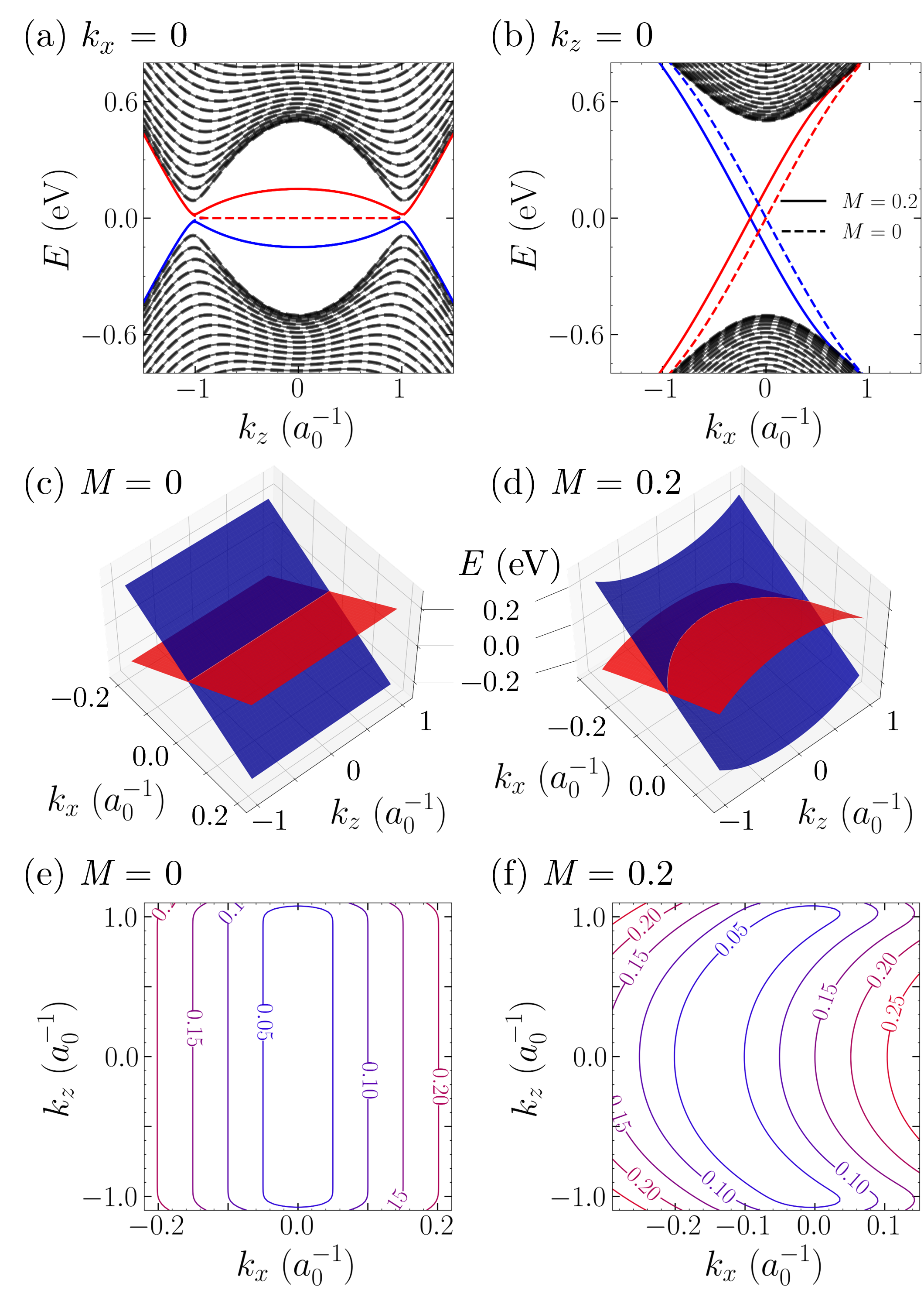}
    \caption{
        Figures are presented to illustrate the energy spectrums of the WSM film with $t_1=t_2=1\,\mathrm{eV}, m=-2.5$ and $N=50$. The two WNs are at approximately $(0, 0, \pm 1)\,a_0^{-1}$.
        The energy spectrums of the chiral surface states (red and blue) and bulk states (black) with $M=0$ (dashed) and $0.2\,\mathrm{eV}$ (solid) at (a) $k_x=0$ and (b) $k_z=0$.
        The energy spectrums of the surface states in a 3D form with (c) $M=0$ and (d) $M=0.2\,\mathrm{eV}$.
        The opposite chiral dispersions are colored in red and blue.
        Panels (e) and (f) display the contour form of the same energy spectrums correspondingly.
        Nonzero $M$ bends the surface states, leading to the asymmetric band structure in the $k_x$ direction.
    }\label{fig1}
\end{figure}

By confining the system in the $y$ direction, we obtain a WSM film with nontrivial upper and lower surfaces, as is shown in Fig.~\ref{fig0}.
Both surfaces hold projections of the WNs connected by Fermi arcs.
Supposing the WSM has $N$ layers in the $y$ direction, the Hamiltonian can be expressed as
\begin{align}
    H_0(\mathbf{k}_\parallel)
        = \sum_{l=1}^{N} \mathcal{H}_0(\mathbf{k}_\parallel) \ket{l}\bra{l}
        + \sum_{l=1}^{N-1} \left(\mathcal{H}_y \ket{l+1}\bra{l}
        + \mathrm{H.c.}\right)
\end{align}
with $\mathcal{H}_0(\mathbf{k}_\parallel) = t_1 \sin (a_0 k_x) \sigma_x \nonumber + t_2 [\cos (a_0 k_x) + \cos (a_0 k_z) + m] \sigma_z$ denoting the onsite terms
and $\mathcal{H}_y = \frac{1}{2} (-i t_1 \sigma_y + t_2 \sigma_z)$ the hopping terms in the $y$ direction,
where $\mathbf{k}_\parallel = (k_x, k_z)$ is the momentum component and $l$ is the layer index.

The energy spectrum of such a WSM film is shown in Figs.~\ref{fig1}(a) and (b) with dashed lines.
In Fig.~\ref{fig1}(a) we can see two Weyl valleys, each possessing an energy gap which gradually closes when $N$ becomes larger.
For sufficiently large $N$, the positions of the two WNs will approach those in the bulk cases.
The two chiral surface states (shown in red and blue) cross the energy gap in an opposite way, linking the upper and lower energy bands, as is shown in Fig.~\ref{fig1}(b).

Unlike the 3D bulk model characterized by Eq.~\eqref{eq.ham_bulk2}, both SIS and TRS are satisfied in the multi-layer model with $M=0$.
Utilizing the \textsc{qsymm} package~\cite{Varjas2018}, we identify the unitary operator related to SIS as $\mathcal{I} = A_N \otimes \sigma_z$, where $A_N$ is the unit antidiagonal square matrix with order $N$,
\begin{align}
    A_N = \sum_{l=1}^{N} \ket{l}\bra{N-l+1}.
\end{align}
The operator transforms the Hamiltonian $H_0$ as
\begin{align}
    \mathcal{I} H_0(\mathbf{k}_\parallel) \mathcal{I}^{-1}
        = H_0(-\mathbf{k}_\parallel).
\end{align}
Since the mirror symmetry along the $z$ direction is always maintained, the mirror symmetry along the $x$ direction is equivalent to SIS, so it will not be further discussed.
TRS is represented by the antiunitary operator as $\mathcal{T} = A_N \otimes \sigma_z \mathcal{K}$, where $\mathcal{K}$ is the complex conjugation operator.
It satisfies
\begin{align}
    \mathcal{T} H_0(\mathbf{k}_\parallel) \mathcal{T}^{-1}
    = H_0(-\mathbf{k}_\parallel),
\end{align}
which is similar to $\mathcal{I}$ as $H_0$ happens to be real.
Besides, the system also satisfies the chiral symmetry (CS), represented by the unitary operator as $\mathcal{C} = A_N \otimes i\sigma_y$ which satisfies
\begin{align}
    \mathcal{C} H_0(\mathbf{k}_\parallel) \mathcal{C}^{-1}
    &= -H_0(\mathbf{k}_\parallel).
\end{align}
As is shown in Fig.~\ref{fig1}, both SIS and TRS guarantee the symmetric band structure along the $k_x$ direction, while CS guarantees the energy spectrum symmetric about $E=0$.

In order to obtain asymmetric bands, potentials $\pm M$ are introduced respectively to the upper ($l=1$) and lower ($l=N$) surfaces, as is shown in Fig.~\ref{fig0}.
The Hamiltonian becomes
\begin{align}
    H_M(\mathbf{k}_\parallel) = H_0(\mathbf{k}_\parallel)
    + M (\ket{1}\bra{1} - \ket{N}\bra{N}).
\end{align}
The corresponding energy spectrum is shown in Figs.~\ref{fig1}(a) and (b) with solid lines.
Although the bulk states barely change, the surface states experience severe deformation because of the surface modification, as is more clearly shown in Figs.~\ref{fig1}(d) and (f).
Compared to the $M=0$ case, nonzero $M$ breaks both SIS and TRS while preserving CS, leading to the asymmetry along the $k_x$ direction.

\subsection{Gauge-invariant theory for nonreciprocal ballistic transport}
Without loss of generality, we first focus on the quantum coherent transport in a 1D mesoscopic system with two terminals.
The electric current $I$ driven by the bias voltages $\mathcal{V}_\mathrm{L,R}$ is expressed as~\cite{Christen1996, Jauho1994, Anantram1995, Datta2005}
\begin{align}\label{eq.I}
    I =-\frac{e}{h}\int \mathrm{d}E\,T\left(E, U\right)
        \left[f_\mathrm{L}(E)-f_\mathrm{R}(E)\right],
\end{align}
where $f_\mathrm{L/R}(E) \equiv f\left(E-\mu_\mathrm{L/R}\right)$ is the Fermi-Dirac distribution function in the left (right) terminal with $\mu_\mathrm{L/R} = E_\mathrm{F} - e\mathcal{V}_\mathrm{L/R}$, $E_\mathrm{F}$ the equilibrium Fermi energy;
$T$ denotes the transmission; the additional term $U$ is the Coulomb potential arising from a finite bias, which satisfies the Poisson equation~\cite{Wang1999}
\begin{align}\label{eq.Poisson}
    \nabla_x^2 U(x) = 4 \pi e n(x)
\end{align}
with $n$ the local charge density.
The potential $U(x)$ plays an essential role for the nonreciprocal ballistic transport as the asymmetric band structures are considered.

Considering the uniformity of our setup in the $x$ direction, the Coulomb potential $U$ remains constant throughout the scattering region.
The local charge density $n$ is thus required to be zero everywhere inside the system, which implies that the local density of electrons remains identical between the equilibrium state $n_0$ and the biased condition $n_U$.
Thus, we obtain
\begin{align}
    n_0 = n_U
    \label{eq.n0=nU}
\end{align}
with
\begin{align}
    n_0 &= \int \mathrm{d}E\,\rho(E, U=0) f_0(E),
    \label{eq.n0}\\
    n_U &= \int \mathrm{d}E\,
    \big[\rho_{\mathrm{L}}(E,U)f_\mathrm{L}(E) + \rho_{\mathrm{R}}(E,U)f_\mathrm{R}(E)\big],
    \label{eq.nU}
\end{align}
where $f_0(E)$ is the zero-bias distribution function, $\rho(E)$ is the local density of states (LDOS), and $\rho_\mathrm{L/R}(E)$ is the injectivity or local partial density of states (LPDOS) of the left (right) terminal~\cite{Buttiker1993,Christen1996, Kramer1996, Wang1997} given by
\begin{align}
    \rho_\mathrm{L/R}(E)
        &= \frac{1}{2\pi} \int \mathrm{d}E'\,\delta(E-E')\sum_n \frac{1} {\hbar |v_{n;\mathrm{L/R}}(E')|}\nonumber\\
        &= \frac{1}{2\pi} \sum_n \frac{1} {\hbar |v_{n;\mathrm{L/R}}(E)|},
        \label{eq.rho}
\end{align}
where $v_{n;\mathrm{L/R}}$ is the velocity corresponding to the incident mode $n$ from the left (right) terminal, $\delta(\cdots)$ is the Dirac delta function.
Here, $\rho_\mathrm{L/R}(E)$ is independent of $x$ in a uniform system.
The role of the constant $U$ can be interpreted as inducing a shift of the energy band described by $\rho_\mathrm{L/R}(E, U) = \rho_\mathrm{L/R}(E+eU)$.
The total injectivity contains the contributions from both terminals as
\begin{align}
    \rho(E) = \rho_\mathrm{L}(E) + \rho_\mathrm{R}(E).
\end{align}

In the limit of zero temperature we obtain
\begin{align}
    \int_{E_\mathrm{F}}^{\mu_\mathrm{L}+eU} \mathrm{d}E\,\rho_{\mathrm{L}}(E)
    + \int_{E_\mathrm{F}}^{\mu_\mathrm{R}+eU} \mathrm{d}E\,\rho_{\mathrm{R}}(E) = 0.
\end{align}
For infinitesimal bias voltages $\mathcal{V}_\mathrm{L,R}$, the Coulomb potential can be calculated to the first order of $\mathcal{V}_\mathrm{L,R}$ as
\begin{align}
    U = u_\mathrm{L} \mathcal{V}_\mathrm{L} + u_\mathrm{R} \mathcal{V}_\mathrm{R} + \cdots
    \label{eq.U}
\end{align}
with characteristic potential~\cite{Christen1996,Wang1999}
\begin{align}
    u_\mathrm{L/R} = \frac{\rho_\mathrm{L/R}(E_\mathrm{F})}{\rho(E_\mathrm{F})}.
    \label{eq.u}
\end{align}
This result is consistent with the local neutral approximation~\cite{Christen1996, Kramer1996, Wang1999}.

To the second order of the infinitesimal bias difference $\mathcal{V} \equiv\mathcal{V}_\mathrm{L}-\mathcal{V}_\mathrm{R}$, Eq.~\eqref{eq.I} reduces to
\begin{align}
    I(\mathcal{V})
        =-\frac{e}{h} \int^{\mu_\mathrm{L}+eU}_{\mu_\mathrm{R}+eU} \mathrm{d}E\,T(E)
            \approx G_1 \mathcal{V} + G_2 \mathcal{V}^2,
    \label{eq.I2}
\end{align}
where the first-order and second-order conductances are expressed as
\begin{align}
    G_1 &= \frac{e^2}{h} T_0,
    \label{eq.G1}\\
    G_2 &= \frac{e^3}{2h} \left(u_\mathrm{L} - u_\mathrm{R}\right) \partial_E T_0,
    \label{eq.G2}
\end{align}
with $T_0 \equiv T(E_\mathrm{F}, U=0)$.
It shows in Eq.~\eqref{eq.I2} that the current solely depends on the bias difference between the two terminals, exhibiting the gauge invariance.

The nonzero second-order conductance $G_2$ manifests the nonreciprocal transport of the system, where the key factor lies in the disparity between $u_\mathrm{L}$ and $u_\mathrm{R}$.
This can be achieved by the asymmetric band structures, with which the opposite propagation states possess unequal velocities, $\left|v_\mathrm{L}(E)\right| \neq \left|v_\mathrm{R}(E)\right|$.
Since the LPDOS is solely determined by the velocities as is derived in Eq.~\eqref{eq.rho}, the asymmetric band structures assign different values to the LPDOS for the two terminals with $\rho_\mathrm{L}(E) \neq \rho_\mathrm{R}(E)$, which further gives $u_\mathrm{L} \neq u_\mathrm{R}$.

While the above analysis focuses on 1D geometry, the essential conclusions are extendable to quasi-2D cases with appropriate modifications.
As our setup is also uniform along the $z$ direction, we simplify our model by assuming the Coulomb potential $U$ to be independent of $z$.
This assumption facilitates a simplified solution for the characteristic potential, analogous to the formulation presented in Eq.~\eqref{eq.u}.
The LPDOS and transmission contain the contribution from all $k_z$ channels and can be expressed as
\begin{align}
    \rho_\mathrm{L/R}(E)
        &= \frac{1}{(2\pi)^2} \int \mathrm{d} k_z\,
            \sum_{n} \frac{\theta\left[\pm v_n^x(k_z, E)\right]}{\left|v_n^x(k_z, E)\right|},\\
    \label{eq.rho2}
    T(E) &= \frac{L_z}{2\pi} \int \mathrm{d} k_z\,\sum_{n} \theta\left[v_n^x(k_z, E)\right],
\end{align}
where $\theta(\cdots)$ is the Heaviside step function, $v_n^x(k_z, E)$ represents velocity of the $k_z$ channel in the left(right) terminal, and $L_z$ denotes the width of the scattering region in the $z$ direction.

\begin{figure*}[t]
    \centering
    \includegraphics[width=2\columnwidth]{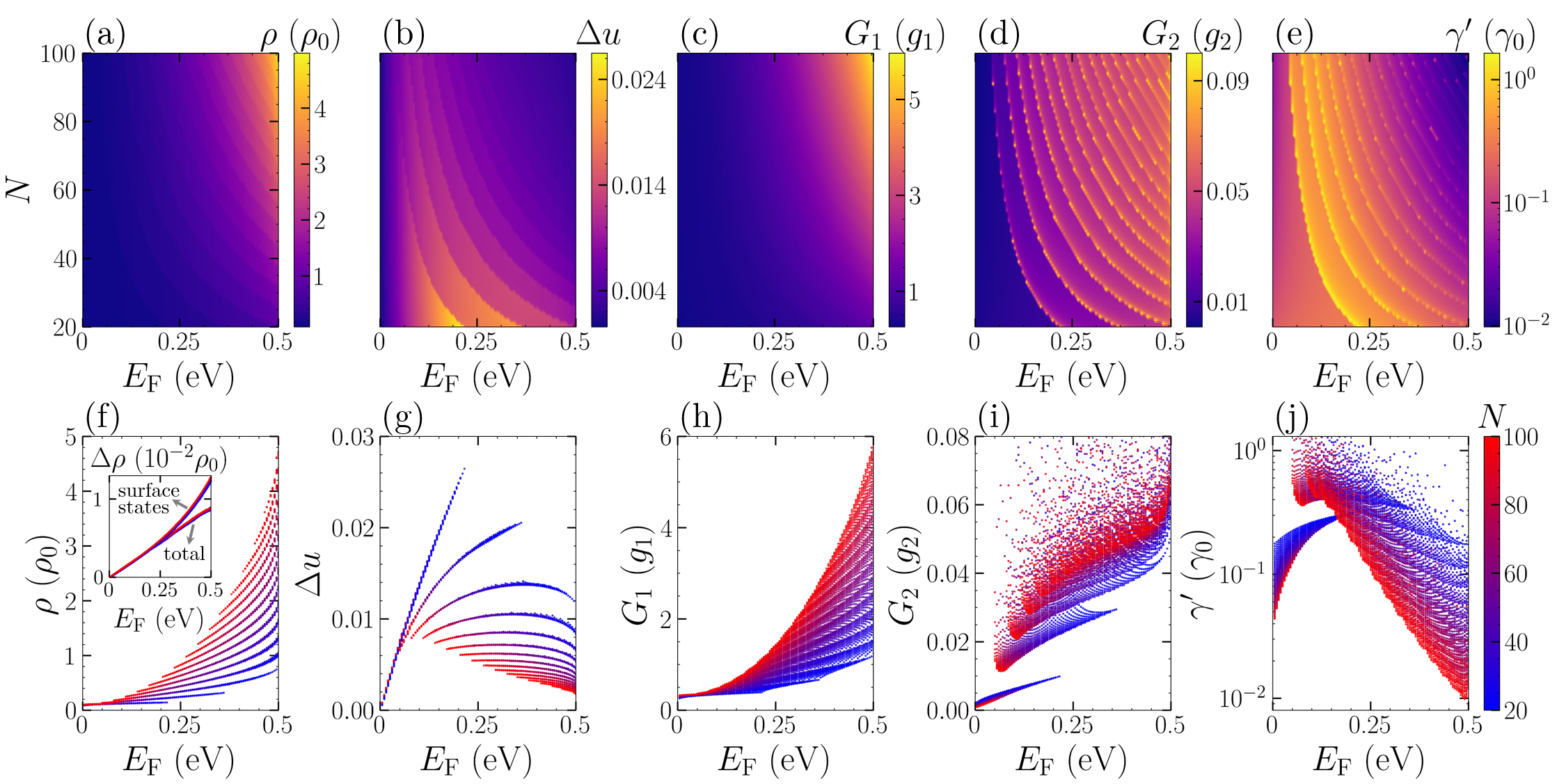}
    \caption{
        Figures are presented to illustrate the dependence of several quantities associated with nonreciprocity on the number of layers $N$ and Fermi energy $E_\mathrm{F}$, with $t_1=t_2=1\,\mathrm{eV}, m=-2.5$ and $M=0.2\,\mathrm{eV}$.
        The upper panels show the contour plots of (a) DOS $\rho$, (b) potential disparity $\Delta u$, (c) first-order conductance $G_1$, (d) second-order conductance $G_2$, and (e) coefficient tensor $\gamma'=\gamma L_z$.
        All contour plots share uniform boundary lines, which correspond to the bottoms of different sub-bands.
        The lower panels (f)-(j) display scatter plots corresponding to the contour plots shown above, with the color scale representing the variation in the number of layers.
        Inset: LPDOS difference $\Delta \rho$, the upper curve representing the contribution from the surface states while the lower representing the total.
        The parameters serving as units are defined such that $\rho_0=\frac{1}{a_0^{2}\,\mathrm{eV}}$, $g_1=\frac{e^2}{h}\frac{L_z}{a_0}$, $g_2=\frac{g_1}{2\,\mathrm{V}}$, and $\gamma_0=\frac{h}{e^2}\frac{a_0}{2\,\mathrm{V}\cdot\mathrm{eV}}\sim 10^{-6}\,\mathrm{m}/(\mathrm{A}\cdot\mathrm{eV})$ with typical $a_0=1\,$\AA.
    }\label{fig2}
\end{figure*}

\section{\label{sec. result} Result and Discussion}

\subsection{Sub-band-resolved nonreciprocal transport signatures}
We start by analyzing the nonreciprocity of the system, which is characterized by the second-order conductance $G_2$ and the coefficient tensor $\gamma$ defined from Eq.~\eqref{eq.I2} as
\begin{align}\label{gamma1}
    R = \frac{\mathcal{V}}{I} = R_0(1-\gamma I M)
\end{align}
with
\begin{align}\label{gamma2}
    R_0 = \frac{1}{G_1}, \quad
    \gamma = \frac{G_2}{G_1^2 M}.
\end{align}
The two quantities, which represent the absolute magnitude and relative strength of the nonreciprocal signal, respectively, are depicted in Figs.~\ref{fig2}(d) and (e).
The two plots are clearly divided into multiple subregions by distinct boundaries, in which the bright yellow indicates pronounced nonreciprocal signals.
The identical division lines, which also appear in Figs.~\ref{fig2}(a)-(c), are formed by Van Hove singularities corresponding to the bottoms of different sub-bands.
Whenever the Fermi energy $E_\mathrm{F}$ crosses the bottom of a sub-band, DOS $\rho$ displays step-like jumps.
Consequently, the potential disparity, defined as $\Delta u \equiv u_\mathrm{L} - u_\mathrm{R}$, exhibits discontinuous variations.
The boundary lines in Fig.~\ref{fig2}(c) are relatively subtle, indicating the continuity of the first-order conductance $G_1$, which is proportional to the transmission $T_0$ as formulated in Eq.~\eqref{eq.G1}.
Although the emergent sub-band makes no contribution to $T_0$, its energy derivative $\partial_E T_0$ changes abruptly, leading to nonzero $G_2$, as is formulated in Eq.~\eqref{eq.G2}.

As is discussed above, the boundary lines correspond to the bottoms of different sub-bands, so the contributions to these subregions originate from the sub-bands lying below the Fermi level.
Both the DOS and transmission increase toward the upper-right region in Figs.~\ref{fig2}(a) and (c), due to the growing number of sub-bands participating in the transport.
The first subregion in the lower-left corresponds to the surface state, and the rest correspond to bulk bands.
In the inset of Fig.~\ref{fig2}(f), we show the LPDOS difference for the two terminals, denoted as $\Delta \rho \equiv \rho_\mathrm{L} - \rho_\mathrm{R}$, in which the lower curve represents the total difference, while the upper shows the contribution from the surface states only.
It reveals that the surface states dominate $\Delta \rho$, the essential origin of nonreciprocity, especially in the low-energy region.
In contrast, the bulk states only contribute a negative correction.
The result is intuitively expected, because the surface engineering induces substantially stronger symmetry breaking to surface states than bulk states, as is manifested in Fig.~\ref{fig1}(b).


The aforementioned physical quantities are presented in the second row of Fig.~\ref{fig2} using scatter plots to facilitate a more intuitive understanding of their respective magnitudes and distribution patterns.
The scatter points aggregate into curve-shaped [Figs.~\ref{fig2}(f) and (g)] and block-shaped [Figs.~\ref{fig2}(h)-(j)] patterns.
Each pattern corresponds to a subregion in the contour plots, revealing the contribution of the relative sub-bands.
For example, the points in the left cluster of Figs.~\ref{fig2}(h)-(j) correspond to the surface states.
Different from other intricate distribution patterns formed by bulk states, they exhibit clear monotonic dependence on both the number of layers and Fermi energy.
Overall, fewer layers and lower energy facilitate the observation of enhanced nonreciprocal effects, with reference to $\gamma$.

\subsection{Single-variable theory of the DOS pattern independent of thickness}
\begin{figure}[t]
    \centering
    \includegraphics[width=1\columnwidth]{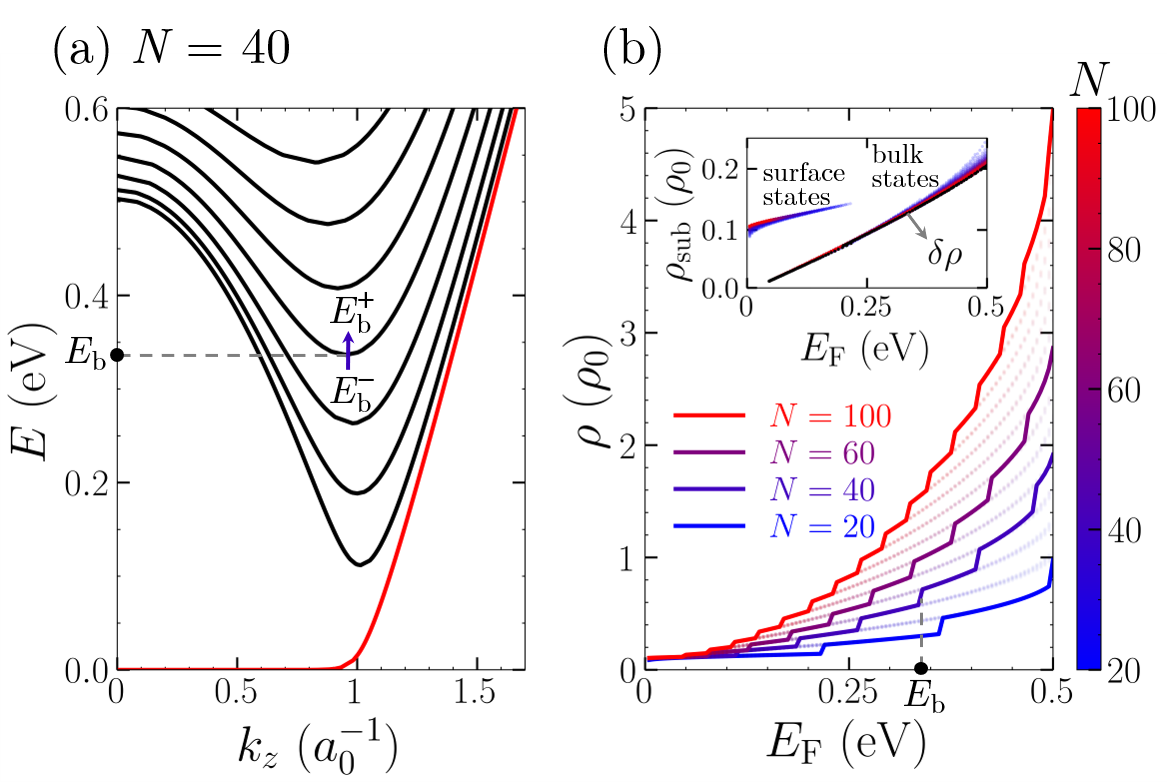}
    \caption{
        (a) Part of the energy dispersion of the WSM film with $N=40$, which shows the sub-band bottoms ($k_x$ not fixed).
        The red line represents the surface state.
        (b) The step-like curves of the DOS as a function of the Fermi energy $E_\mathrm{F}$ with $N=20, 40, 60$ and $100$.
        The light-colored background is the same as Fig.~\ref{fig2}(f).
        Inset: the variation of $\delta\rho$ (black dots) with respect to $E_\mathrm{F}$ according to Eq.~\eqref{eq.deltarho2}, as well as part of the DOS of each single sub-band (colored dots).
        Both obtained in systems with $20 \le N \le 100$.
        The colored dots cluster into two separate curve-shaped patterns, contributed by the surfaces states and bulk states, respectively.
        The black dots and colored dots of bulk states show great consistency, revealing the rough independence of $N$.
        The parameters are the same as those in Fig.~\ref{fig2}.
    }\label{fig3}
\end{figure}

Remarkably, the points representing DOS and potential disparity form curve-like clusters in Figs.~\ref{fig2}(f) and (g).
In this part, we will try to explain how this special pattern is formed.
In Fig.~\ref{fig3}(b), the step-like curves of DOS $\rho$ for several numbers of layers $N$ are shown as a function of Fermi energy $E_\mathrm{F}$.
For a given $N$, when $E_\mathrm{F}$ crosses the bottom of a sub-band, $E_\mathrm{b}$ for example, DOS $\rho$ undergoes a jump to the step of the next index.
All the steps sharing the same index from different $N$ merge into a curve, representing the contribution of corresponding sub-bands.

We shall prove that such DOS jump depends solely on energy and is independent of number of layers.
First, we assume that a sub-band is one-to-one correspondent to its band bottom energy within a certain energy valley (excluding the surface states).
Second, in systems with varying numbers of layers $N$, sub-bands sharing the same bottom energy are assumed to exhibit approximately identical dispersion relations, provided that $N$ is not too small.
With these two assumptions, a sub-band within a single valley can be labeled by its band bottom energy, regardless of the number of layers.

Now we suppose that a sub-band has the bottom energy of $E_\mathrm{b}$, with the corresponding wave vector $\mathbf{k}_{\parallel\mathrm{b}} = (k_{x\mathrm{b}}, k_{z\mathrm{b}})$.
Expanding the energy spectrum near the band bottom to the second order using Taylor expansion, we get
\begin{align}
    E_{\mathbf{k}_\parallel} = E_\mathrm{b}
        + \alpha_{xx}(k_x-k_{x\mathrm{b}})^2
        + \alpha_{zz}(k_z-k_{z\mathrm{b}})^2,
    \label{eq.ek_expansion}
\end{align}
where $\alpha_{xx,zz}$ are the second-order coefficients of the expansion.
Since $E_\mathrm{b}$ is the minimum energy of the sub-band, the first-order coefficients $\alpha_{x,z}$ vanish and $\alpha_{xx,zz}$ are positive.
With the residue theorem, the DOS of this sub-band can be calculated as
\begin{align}
    \rho(E; E_\mathrm{b})
        &= \frac{1}{(2\pi)^2}\int\mathrm{d}\mathbf{k}_\parallel\,\delta(E-E_{\mathbf{k}_\parallel})\nonumber\\
        &= \frac{1}{4\pi\sqrt{\alpha_{xx}(E_\mathrm{b})\alpha_{zz}(E_\mathrm{b})}}
    \theta(E-E_\mathrm{b}).
\end{align}
It is noteworthy that different sub-bands exhibit distinct dispersion relations, so generally $\alpha_{xx,zz}$ is dependent on $E_\mathrm{b}$.
Then we can define the DOS jump at the bottom energy of this sub-band as
\begin{align}
    \delta\rho(E_\mathrm{b}) &\equiv
        \rho(E_\mathrm{b}^+; E_\mathrm{b}) - \rho(E_\mathrm{b}^-; E_\mathrm{b})\nonumber\\
        &= \frac{1}{4\pi\sqrt{\alpha_{xx}(E_\mathrm{b})\alpha_{zz}(E_\mathrm{b})}},
    \label{eq.deltarho}
\end{align}
in which $\rho(E_\mathrm{b}^\pm; E_\mathrm{b})$ is the limit of $\rho(E; E_\mathrm{b})$ as $E$ approaches $E_\mathrm{b}$ from above or below, as is shown in Fig.~\ref{fig3}(a).
Since other sub-bands with different bottom energy from $E_\mathrm{b}$ have no contribute to the DOS jump at $E_\mathrm{b}$, so $\delta\rho(E_\mathrm{b})$ is the total DOS jump.

For multi-valley systems, the contribution of each single valley should be taken into account.
With valleys labeled by $\nu$, a more comprehensive formula of Eq.~\eqref{eq.deltarho} can be written as
\begin{align}
    \delta\rho(E) &= \sum_\nu \delta\rho^\nu(E)
        = \sum_\nu \frac{1}{4\pi\sqrt{\alpha^\nu_{xx}(E)\alpha^\nu_{zz}(E)}}.
    \label{eq.deltarho1}
\end{align}
Considering the two-node WSM systems have two symmetrical valleys, the total DOS jump is
\begin{align}
    \delta\rho(E) = \frac{1}{2\pi\sqrt{\alpha_{xx}(E)\alpha_{zz}(E)}}.
    \label{eq.deltarho2}
\end{align}

Eqs.~\eqref{eq.deltarho1} and \eqref{eq.deltarho2} reveal that the DOS jump $\delta\rho$ is a single-variable function of $E$, which is independent of $N$.
To demonstrate this, we plot the variation of $\delta\rho$ with respect to the Fermi energy $E_\mathrm{F}$ in black dots in the inset of Fig.~\ref{fig3}(b), according to Eq.~\eqref{eq.deltarho2} using $\alpha_{xx,zz}$ from the sub-bands in systems with $20 \le N \le 100$.
In the same inset, we also show the DOS contributed by each single sub-band, denoted as $\rho_{\mathrm{sub}}$, in colored dots, where the part of bulk states corresponds to the difference of DOS between the adjacent curves.
The black dots and colored dots of bulk states are distributed on the same curve-shaped pattern, where the variation in $N$ leads to little deviation, clearly demonstrating the independence of $N$.

\section{\label{sec. conclusion} Conclusion}
In conclusion, we research the nonreciprocal ballistic transport in multi-layer WSM films with surface engineering using the gauge-invariant theory.
We find that the nonreciprocity comes almost entirely from the surface states while the bulk states barely have contribution.
We also find that increasing thickness of the WSM film will suppress the nonreciprocity, which may be helpful for experimental discovery of significant nonreciprocal signals in the ballistic regime.
Finally, we put forward a single-variable theory to explain the pattern of the DOS which is roughly independent of thickness in multi-layer systems.
We hope it to spark inspirations for other researchers.

\begin{acknowledgments}
This work was supported by the National Natural Science Foundation of China under Grant No. 12304068 (H.G.), No. 12274235 (R.M.), No. 12174182 (D.Y.X.), the startup Fund of Nanjing University of Aeronautics and Astronautics Grant No. YAH24076 (H.G.), and the State Key Program for Basic Researches of China under Grants No. 2021YFA1400403 (D.Y.X.). 
The computations are partially supported by High Performance Computing Platform of Nanjing University of Aeronautics and Astronautics.
\end{acknowledgments}

\appendix

\section{Non-equilibrium Green's function method for nonuniform systems}
The non-equilibrium Green's function method is particularly suited for systems nonuniform in the $x$ direction.
The transmission in Eq.~\eqref{eq.I} is given by $T = \mathrm{Tr} \left(\Gamma_\mathrm{L}\mathcal{G} \Gamma_\mathrm{R}\mathcal{G}^{\dagger}\right)$, where $\mathcal{G}$ is the retarded Green's function defined as
\begin{align}
    \mathcal{G}(E,U) = \left[E - H + eU
        -(\Sigma_\mathrm{L} + \Sigma_\mathrm{R})\right]^{-1}
    \label{eq.G}
\end{align}
with $\Sigma_\mathrm{L/R}$ the self-energy introduced by the left (right) terminal.
The linewidth functions are defined as $\Gamma_\mathrm{L/R} = i\left(\Sigma_\mathrm{L/R} - \Sigma_\mathrm{L/R}^\dagger\right)$.

In Eq.~\eqref{eq.G}, $H$ is the Hamiltonian of the system in equilibrium.
The Coulomb potential $U(x)$ satisfies the Poisson equation~\cite{Wang1999}
\begin{align}
    \nabla^2 U(x) = 4 \pi i e \int\frac{\mathrm{d}E}{2\pi}\,\left[\mathcal{G}^{<}(E, U)\right]_{xx},
    \label{eq.Poisson2}
\end{align}
where $x$ denotes the position.
The lesser Green's function $\mathcal{G}^{<}$ is defined as $\mathcal{G}^{<} = \mathcal{G} \Sigma^{<} \mathcal{G}^{\dagger}$ with $\Sigma^{<} = i \left(\Gamma_\mathrm{L} f_\mathrm{L} + \Gamma_\mathrm{R} f_\mathrm{R}\right)$.

In general, a self-consistent approach is required to solve Eqs.~\eqref{eq.G} and \eqref{eq.Poisson2} in the nonlinear regime.
Since the lesser Green's function exhibits a nonlinear relationship with $U$, Eq.~\eqref{eq.Poisson2} is a nonlinear differential equation.
Here, we focus on the weakly nonlinear regime, where the Coulomb potential can be expanded to the first order of $\mathcal{V}_\mathrm{L/R}$ as
\begin{align}\label{Uexpandeq}
    U(x) = u_\mathrm{L}(x) \mathcal{V}_\mathrm{L} 
        + u_\mathrm{R}(x) \mathcal{V}_\mathrm{R} + \cdots,
\end{align}
where the zeroth-order term (potential in equilibrium) has been absorbed into
the Hamiltonian $H$, and $u_\mathrm{L/R}(x)$ denotes the characteristic potential.
Gauge invariance of the theory requires~\cite{Christen1996}
\begin{align}
    u_\mathrm{L} + u_\mathrm{R} = 1.
\end{align}
To the lowest order, we derive the equation for $u_\mathrm{L/R}(x)$ from Eqs.~\eqref{eq.Poisson2} and \eqref{Uexpandeq}~\cite{Buttiker1993,Christen1996} as
\begin{align}\label{cpueq}
    - \nabla^2_x u_\mathrm{L/R} + 4 \pi e^2 \rho\,u_\mathrm{L/R}
    = 4\pi e^2 \rho_\mathrm{L/R},
\end{align}
where the LDOS $\rho(x)$ and the LPDOS $\rho_\mathrm{L/R}(x)$ are given by \cite{Buttiker1993,Christen1996}
\begin{align}
    \rho(x) &= \rho_\mathrm{L}(x) + \rho_\mathrm{R}(x),\\
    \rho_\mathrm{L/R}(x)
        &= \frac{1}{2\pi} \int \mathrm{d}E\,\left(-\partial_E f_0\right)
        \left( \mathcal{G}_0\Gamma_{\alpha}
            \mathcal{G}_0^\dagger \right)_{xx},
\end{align}
in which $\mathcal{G}_0$ is the equilibrium Green's function with $U(x)=0$.
An alternative expression for injectivity in terms of scattering wavefunctions and the velocity of incident modes is~\cite{Kramer1996, Wang1997}
\begin{align}\label{LPDOSPSiEq}
    \rho_\mathrm{L/R}(x)
        = \frac{1}{2\pi} \int \mathrm{d}E\,\left(-\partial_E f_0\right)
            \sum_n \frac{|\Psi_{n;\mathrm{L/R}}(x)|^2} {\hbar |v_{n;\mathrm{L/R}}|},
\end{align}
where $\Psi_{n;\mathrm{L/R}}(x)$ and $v_{n;\mathrm{L/R}}$ are the scattering wave function and velocity corresponding to the incident mode $n$ from the left (right) terminal.

For a uniform system, the spatial distribution of the scattering wavefunctions $\Psi_{n;\mathrm{L/R}}(x)$ is independent of $x$, and so is the LPDOS $\rho_\mathrm{L/R}(x)$.
Similarly, the Coulomb potential remains constant throughout the scattering region so that the characteristic potentials satisfy $\nabla_x^2 u_\mathrm{L/R} = 0$.
The characteristic potentials are then expressed as
\begin{align}\label{cpuSoleq}
    u_\mathrm{L/R} = \frac{\rho_\mathrm{L/R}}{\rho},
\end{align}
which recovers the results in Eqs.~\eqref{eq.rho} and \eqref{eq.u}.

\begin{figure}[t]
    \centering
    \includegraphics[width=1\columnwidth]{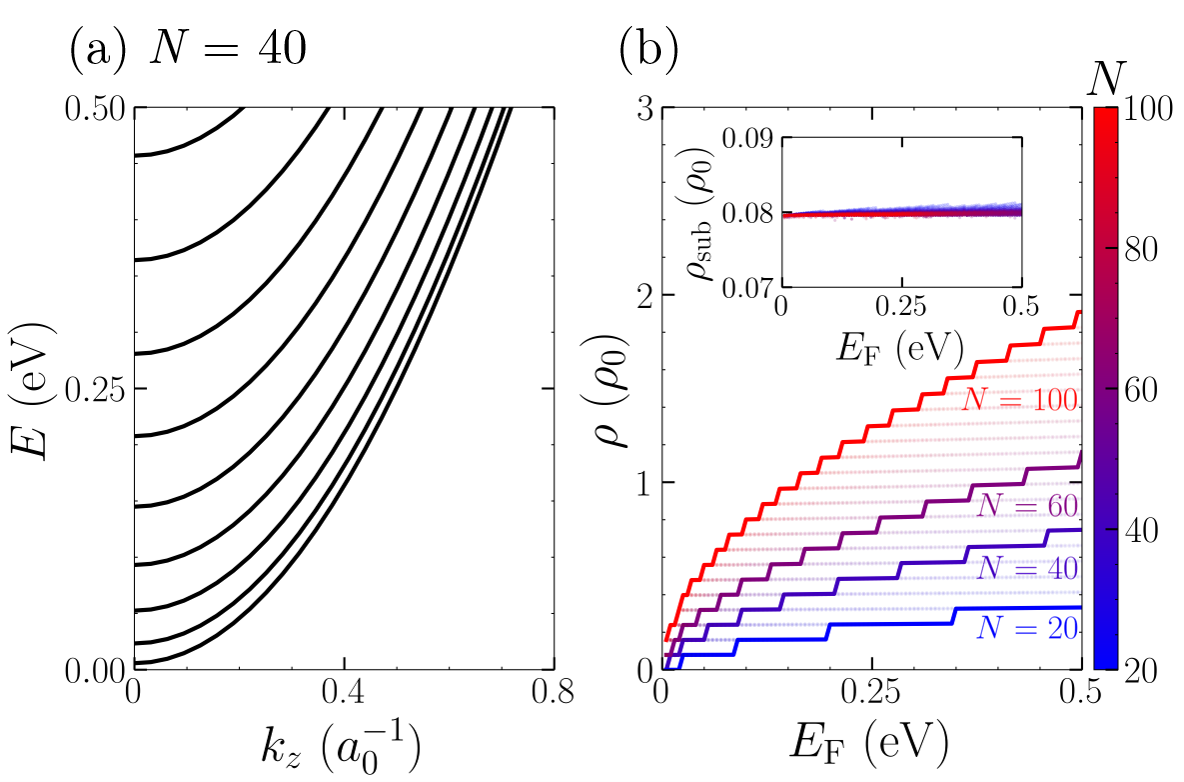}
    \caption{
        (a) Part of the energy dispersion of the NM film with $N=40$, showing the sub-band bottoms with $k_x=0$.
        (b) The step-like curves of the DOS as a function of the Fermi energy $E_\mathrm{F}$ with $N=20, 40, 60$ and $100$.
        The light-colored background is the scatter plot of DOS with $20 \le N \le 100$.
        Inset: part of the DOS of each single sub-band.
        The horizontal distribution shows the rough independence of both $E_\mathrm{F}$ and $N$.
    }\label{fig4}
\end{figure}

\section{DOS pattern for Normal Metal films}
Consider the following Hamiltonian of the multi-layer normal metal (NM) films,
\begin{align}
    H_0(\mathbf{k}_\parallel)
        = \sum_{l=1}^{N} \mathcal{H}_0(\mathbf{k}_\parallel) \ket{l}\bra{l}
        + \sum_{l=1}^{N-1} \left(\mathcal{H}_y \ket{l+1}\bra{l}
        + \mathrm{H.c.}\right)
\end{align}
with $\mathcal{H}_0(\mathbf{k}_\parallel) = -2 [\cos (a_0 k_x) + \cos (a_0 k_z) - 3]\,\mathrm{eV}$ and $\mathcal{H}_y=1\,\mathrm{eV}$.
As is shown in Fig.~\ref{fig4}(a), the energy spectrum of each sub-band is paraboloid-like, which can be roughly described by Eq.~\eqref{eq.ek_expansion} with $k_{x\mathrm{b}} = k_{z\mathrm{b}} = 0$ and $\alpha_{xx} = \alpha_{zz} = \rho_0^{-1}$.
Because there is only a single valley, the DOS jump can be calculated according to Eq.~\eqref{eq.deltarho} as
\begin{align}
    \delta\rho(E) = \frac{1}{4\pi\rho_0^{-1}} \approx 8.0\times 10^{-2} \,\rho_0,
\end{align}
which is a constant independent of energy, consistent with the inset in Fig.~\ref{fig4}(b).
This can attribute to the identity of all sub-bands.

The shapes of the envelope of Figs.~\ref{fig3}(b) and \ref{fig4}(b) are also consistent with the DOS of their bulk materials,
\begin{align}
    \rho^{\mathrm{WSM}}_{\mathrm{bulk}}(E)
        &\sim \frac{1}{(2\pi)^3}\int \mathrm{d}\mathbf{k}\,\delta(E-k) \propto E^2,\\
    \rho^{\mathrm{NM}}_{\mathrm{bulk}}(E)
        &\sim \frac{1}{(2\pi)^3}\int \mathrm{d}\mathbf{k}\,\delta(E-k^2)\propto E^{1/2},
\end{align}
which are respectively roughly consistent with the red curves in Figs.~\ref{fig3}(b) and \ref{fig4}(b).

Similar discussion can be found in Refs.~\cite{Aydin2016} and \cite{Aydin2019}.

\bibliography{main}

\end{document}